\documentclass[11pt,ams]{article}
\usepackage[latin9]{inputenc}
\usepackage{amsmath}
\usepackage{amssymb}
\usepackage{esint}

\makeatletter

\providecommand{\tabularnewline}{\\}

\usepackage{amsfonts} 
\makeatother

\begin{document}

\title{\textbf{Landau confining replica model from an explicitly breaking
of a $SU(3)$ group without auxiliary fields.}}

\author{\textbf{M.~M.~Amaral}%
\thanks{macielamaral@uerj.br%
}\,\,, \textbf{M. A. L. Capri}%
\thanks{caprimarcio@gmail.com%
}\,\,, \textbf{Y.~E.~Chifarelli}%
\thanks{yveseduardo@gmail.com%
}\,\,, \textbf{V.~E.~R.~Lemes}%
\thanks{vitor@dft.if.uerj.br%
}\,\,,\\[2mm] \textit{\small{{UERJ $-$ Universidade do Estado
do Rio de Janeiro}}}\\
 {\small{{} }}\textit{\small{{Instituto de Física $-$ Departamento
de Física Teórica}}}\\
 {\small{{} }}\textit{\small{{Rua S{ã}o Francisco Xavier 524,
20550-013 Maracan{ã}, Rio de Janeiro, RJ, Brasil.}}}}

\maketitle
\vspace{-1cm}

\begin{abstract}
We propose a mechanism displaying gluon confinement, as defined by
the behavior of the propagators, in a model of $SU(2)$ gauge fields.
The model originates from an explicitly broken $SU(3)$ gauge theory
giving rise to a replica model composed of three mixed $SU(2)$ groups.
The mechanism consists in the usual $SU(3)$ Yang-Mills theory in
the Landau gauge, with a soft breaking term in such a way as to change
the field propagation and group content at low energies. The relation
of this soft mass term with the Gribov problem is presented and the
link between soft terms and the scaling and decoupling solutions is
discussed.


\end{abstract}
\setcounter{page}{0}\thispagestyle{empty}

\vfill{}
 \newpage{}\ \makeatother


\section{Introduction}


The problem of gluon confinement is the core of the general investigations
of strongly coupled gauge theories. An important aspect of gluon confinement
is related to the behavior of the gluon propagator.

Recently this problem has received great attention in different approaches.
One of these approaches comes from lattice simulation where the behavior
of the gluon propagator in the infrared regime is studied \cite{Cucchieri:2007md,Cucchieri:2008fc,Bogolubsky:2007ud}.
These results display positivity violation thus making impossible
a particle interpretation for the gluon excitation at low energies.
This is taken as a strong signal of gluon confinement. In the analytical
point of view, one possible approach of the confinement problem comes
from the analysis of the Gribov copies \cite{Gribov:1977wm}, where
the Gribov-Zwanziger (GZ) model \cite{Zwanziger:1991ac,Zwanziger:1992qr},
and this refined version, the so-called Refined Gribov-Zwanziger (RGZ)
model \cite{Dudal:2008sp}, take place. Also, a recently developed
model based on the introduction of a replica of the Faddeev-Popov
action enjoys a confined gluon propagator%
\footnote{We will refer to this model just as \textit{the replica model}.%
} \cite{Sorella:2010it}. Usually, these models provide propagators
behaving as%
\footnote{In the case of the RGZ model the gluon propagator take into account
the effects of dimension two condensates and it is more complex. Typically,
the gluon form factor looks like

\begin{center}
$D(p{2})=\frac{p{2}+M{2}}{p{4}+(M{2}+m{2})p{2}+(\gamma{4}+M{2}m{2})},$ 
\par\end{center}

where $M$ and $m$ are mass parameters associated with the condensates,
while $\gamma$ is the so-called Gribov mass parameter.%
} 
\begin{equation}
D(p^{2})=\frac{p^{2}}{p^{4}+\gamma^{4}}\,,\label{Gribov_like}
\end{equation}
where $D(p^{2})$ is the gluon form factor in Euclidean spacetime
and $\gamma$ is a mass parameter%
\footnote{In the GZ model this parameter is known as the Gribov parameter, which
is directly associated with the restriction of the Feynman path integrals
to the Gribov region.%
}. It is easy to notice that a propagator of this type has complex
poles, being impossible its identification with a propagation of a
physical particle. In other words, it is a suitable candidate to be
a confined object.\\
 \\
 An important feature that has been studied about propagators like
(\ref{Gribov_like}) is that they can be seen as a propagation of
two unphysical modes with imaginary squared masses $\pm i\gamma^{2}$,
named $i$-particles \cite{Baulieu:2009ha}. In fact, one can immediately
notice that the form factor (\ref{Gribov_like}) can be written as
\begin{equation}
\frac{p^{2}}{p^{4}+\gamma^{4}}=\frac{1}{2}\left(\frac{1}{p^{2}-i\gamma^{2}}+\frac{1}{p^{2}+i\gamma^{2}}\right)\,.\label{i-propagator}
\end{equation}
Despite the fact that such propagators do not have an interpretation
in the physical spectrum, it is still possible to construct composite
operators, $\mathcal{O}[A]$, whose the correlation functions exhibit
a Källén-Lehmann spectral representation: 
\begin{equation}
\langle\mathcal{O}(p)\mathcal{O}(-p)\rangle=\int_{\tau_{0}}^{\infty}d\tau\,\frac{\rho(\tau)}{\tau+p^{2}}\,,\label{KL}
\end{equation}
where $\rho(\tau)$ is the positive spectral density and $\tau_{0}\geq0$
stands for the threshold. An important feature of expression (\ref{KL})
is that we can move from Euclidean to Minkowski space. Moreover, the
positivity of the spectral density $\rho(\tau)$ enables us to give
an interpretation of (\ref{KL}) in terms of physical states with
positive norm.\\
 \\
 In the present work, we will study a $SU(3)$ Euclidean Yang-Mills
theory and try to explore a well known property of this group, which
consists in the fact that it contains three mixed $SU(2)$ groups
\cite{Kaku:1993ym,Narison:2002pw} . In the presence of a dimension
two condensate, $\langle d^{abc}A_{\mu}^{b}A_{\mu}^{c}\rangle$, with
$d^{abc}$ being the totally symmetric structure constant of the $SU(3)$
Lie algebra, the color group symmetry is explicitly broken. However,
two of these three $SU(2)$ groups are related in a structure of $i$-particles,
and, thanks to a residual symmetry, it is possible to write composite
operators having a spectral representation like (\ref{KL}). \\
 \\
 The paper is organized as follows. In Sect.~$2$, we make a brief
review about the GZ model, the introduction of the $i$-particles,
and the replica model. In Sect.~$3$, we introduce the condensate
$\langle d^{abc}A_{\mu}^{b}A_{\mu}^{c}\rangle$ and write the action
corresponding to the model that we desire to study. Also in this section
we show how this condensate modifies the gluon propagator and how
the $i$-particle structure appears between two of the three $SU(2)$
groups. In Sect. $4$, the relation of the mass terms, obtained from
the introduction of dimension two condensates, and the scaling and
decoupling solution is established. The Sect.~$5$ is dedicated to
discuss the Källén-Lehmann spectral representation of the candidates
to be a physical observable. Our conclusions are presented in Sect.~$6$.
The symmetry content of the model, characterized by a full set of
Ward identities compatible with the Quantum Principle Action \cite{Piguet:1995er},
and the proof of its renormalizability are presented in the Appendix
A. And some properties of $SU(3)$ groups are presented in Appendix
B.


\section{A brief review: the GZ model, $i$-particles and the replica model}



\subsection{The GZ model}

In order to clarify the understanding of the Gribov problem and its
correlation with our proposal we will present below a description
of the problem in the Landau gauge. The Euclidean $SU(N)$ Yang-Mills
action in the Landau gauge is given by: 
\begin{equation}
S_{\mathrm{YM}}=\int d^{4}x\,\left(\frac{1}{4}F_{\mu\nu}^{a}F_{\mu\nu}^{a}+ib^{a}\,\partial_{\mu}A_{\mu}^{a}+\overline{c}^{a}\partial_{\mu}D_{\mu}^{ab}c^{b}\right)\,,\label{YM}
\end{equation}
where 
\begin{equation}
F_{\mu\nu}^{a}=\partial_{\mu}A_{\nu}^{a}-\partial_{\nu}A_{\mu}^{a}+gf^{abc}A_{\mu}^{b}A_{\nu}^{c}
\end{equation}
and 
\begin{equation}
D_{\mu}^{ab}=\delta^{ab}\partial_{\mu}-gf^{abc}A_{\mu}^{c}\,.
\end{equation}
Here, $A_{\mu}^{a}$ is the gauge field, $b^{a}$ is a Lagrange multiplier
enforcing the Landau gauge, $\partial_{\mu}A_{\mu}^{a}=0$, $(\bar{c}^{a},c^{a})$
are a pair of anti-commuting scalar fields known as the Faddeev-Popov
ghost fields, and $g$ is the coupling constant of the theory. The
labels $(a,b,c,\dots)$ run to $1$ to $(N^{2}-1)$ and $f^{abc}$
are the totally anti-symmetric structure constant of the Lie algebra
of the generator of $SU(N)$. Also, this action is left invariant
under the following nilpotent BRST transformations: 
\begin{equation}
sA_{\mu}^{a}=-D_{\mu}^{ab}c^{b},\hspace{0.5cm}sc^{a}=\frac{g}{2}f^{abc}c^{b}c^{c},\hspace{0.5cm}s\overline{c}^{a}=ib^{a},\hspace{0.5cm}sb^{a}=0.\label{brs-ym}
\end{equation}
Although the gauge be fixed by the Faddeev-Popov method, Gribov showed
in \cite{Gribov:1977wm}%
\footnote{See also \cite{Sobreiro:2005ec} for a pedagogical review.%
} that there are still field configurations obeying the Landau gauge
linked by gauge transformations, \textit{i.e.} there are still equivalent
configurations, or copies, being taken into account into the Feynman
path integral. In other words, the gauge is not completely fixed and
the remaining ambiguity is allowed due to the existence of normalizable
zero-modes of the Faddeev-Popov operator, 
\begin{equation}
\mathcal{M}^{ab}=-\partial_{\mu}D_{\mu}^{ab}\,.
\end{equation}
Gribov also showed that to eliminate these copies the domain of integration
of the functional integral should be restricted to a certain region
$\Omega$, the so-called Gribov region, that is defined as the set
of field configurations performing the Landau gauge condition, for
which the Faddeev-Popov operator is strictly positive, namely 
\begin{equation}
\Omega:=\{\, A_{\mu}^{a}\,|\,\partial_{\mu}A_{\mu}^{a}=0,\,\mathcal{M}^{ab}(A)>0\,\}\,.\label{definicao}
\end{equation}
Its boundary, $\partial\Omega$, where the first vanishing eigenvalue
of the Faddeev-Popov operator shows up, is known as the Gribov horizon.\\
 \\
 As in the region $\Omega$ the Faddeev-Popov operator is positive
than its inverse must diverge when approaching the horizon, due to
the existence of a zero mode. So the restriction to the first Gribov
region is implemented requiring that 
\begin{equation}
G(p^{2},A)=\frac{\delta^{ab}}{N^{2}-1}\langle p|(-\partial_{\mu}D_{\mu}^{ab})^{-1}|p\rangle\,,
\end{equation}
which is the normalized trace of the ghost connected two point function
in momentum space, has no pole for a given nonvanishing value of the
momentum $p$, except for the singularity at $p=0$, corresponding
to the first Gribov horizon. At $p\approx0$ one can write 
\begin{eqnarray}
G(p^{2},A) & \approx & \frac{1}{p^{2}}\frac{1}{1-\sigma(p^{2},A)}\,,\label{eq:twopointghosparam}\\
\sigma(p^{2},A) & = & \frac{N}{N^{2}-1}\frac{1}{p^{2}}\int\frac{d^{4}q}{(2\pi)^{4}}\frac{(p-q)_{\mu}p_{\nu}}{(p-q)^{2}}A_{\mu}^{a}(-q)A_{\nu}^{a}(q).\label{defregion}
\end{eqnarray}
From the above expression (\ref{defregion}), it follows that the
no-pole condition at finite nonvanishing $p$ is 
\begin{equation}
\sigma(p^{2},A)<1.\label{eq:nopolecondictiongribov}
\end{equation}
As $\sigma(p^{2},A)$ decreases as $p^{2}$ increases one can also
take 
\begin{equation}
\sigma(0,A)=\frac{1}{4}\frac{N}{N^{2}-1}\int\frac{d^{4}q}{(2\pi)^{4}}\frac{1}{q^{2}}(A_{\mu}^{a}(-q)A_{\mu}^{a}(q))\leq1\,.\label{sigma}
\end{equation}
It is important to emphasize here that we work with the trace of the
ghost propagator to find the restriction to the Gribov region. This
is a particularity of the Gribov mechanism in the Landau gauge. This
is related to the convexity of the Gribov region in the Landau gauge.
Other gauges, like the maximal Abelian gauge, does not necessarily
present the same property \cite{Capri:2008vk,Capri:2010an,Capri:2005tj}%
\footnote{In the maximal Abelian gauge we take only the trace of diagonal ghost
propagator.%
}.

In order to perform the restriction to the Gribov region into the
partition function, $\mathcal{Z}$, the final step is to introduce
the no-pole condition with the help of a Heaviside function: 
\begin{equation}
{\cal {Z}}=\int{\cal {D}}A\delta(\partial A)\theta(1-\sigma(0,A))\exp^{-S_{\mathrm{YM}}}.
\end{equation}
This will give rise to a propagator for the gauge field of the type
\begin{equation}
\langle A_{\mu}^{a}(-q)A_{\nu}^{b}(q)\rangle=\delta^{ab}\frac{q^{2}}{q^{4}+\gamma^{4}}\left(\delta_{\mu\nu}-\frac{q_{\mu}q_{\nu}}{q^{2}}\right)\,,
\end{equation}

Note that the only allowed singularity at (\ref{eq:twopointghosparam})
is at $p^{2}=0$, whose meaning is that of approaching the horizon,
where $G(p^{2},A)$ is singular due to the appearance of zero modes
of the Faddeev-Popov operator. Thus we have to take \cite{Gribov:1977wm}:

\begin{equation}
\sigma(0,A)=1\label{eq:sigma=00003D00003D00003D1}
\end{equation}

And thus the Gribov parameter $\gamma$ is fixed by the gap equation:
\begin{equation}
\frac{3Ng^{2}}{4}\int\frac{d^{4}q}{(2\pi)^{4}}\frac{1}{q^{4}+\gamma^{4}}=1.\label{gap}
\end{equation}
It is clear that the Gribov approach is only the first step in order
to consistently treat the problem of zero modes and the Gribov copies
in a gauge fixed Yang-Mills theory. The second step is the GZ theory
\cite{Zwanziger:1991ac,Zwanziger:1992qr}, which consists in a renormalizable
and local way to implement the restriction to the first Gribov region.
In fact, Zwanziger observed that the restriction could be implemented
by adding the following term in the action (\ref{YM}): 
\begin{equation}
S_{\mathrm{GZ}}=S_{\mathrm{YM}}+\gamma^{4}H(A)\,,
\end{equation}
where, $H(A)$ is the so-called horizon function, 
\begin{equation}
H(A)=g^{2}\int d^{4}x\, d^{4}y\, f^{abc}A_{\mu}^{b}(x)[\mathcal{M}^{-1}]^{ad}(x,y)f^{dec}A_{\mu}^{e}\,.
\end{equation}
In the Zwanziger approach, the parameter $\gamma$ is fixed by the
equation 
\begin{equation}
\langle H(A)\rangle=4V(N^{2}-1)\,,\label{gapGZ}
\end{equation}
where $V$ is the Euclidean space volume. Notice that the Gribov form
factor (\ref{sigma}) coincides with the first order of the horizon
function%
\footnote{Actually, the form factor can be calculated at all orders and the
result is that such coincidence occurs in fact at all orders. In \cite{Gomez:2009tj},
this equivalence is proved at third order, and in \cite{Guimaraes}
it is proved at all orders.%
}: 
\begin{equation}
\frac{H(A)}{4V(N^{2}-1)}=\sigma(0,A)+O(A^{3})\,.
\end{equation}
It is clear that the horizon function is nonlocal, but it can be localized
with the help of a suitable set of auxiliary fields. In order to ensure
that those extra fields do not introduce extra degrees of freedom
they are introduced in the form of a BRST quartet%
\footnote{Actually, the BRST quartet is composed by two BRST doublets, which
has the basic structure

\begin{center}
sU=V\,,\quad{}sV=0\,, 
\par\end{center}

for a generic pair of fields $(U,V)$, guarantying the nilpotence
of the BRST operator, $s$, \textit{i.e.} $s^{2}=0$. %
}: 
\begin{eqnarray}
 &  & s{\bar{\omega}}_{\mu}^{ab}={\bar{\varphi}}_{\mu}^{ab}\,,\qquad s{\bar{\varphi}}_{\mu}^{ab}=0\,,\nonumber \\
 &  & s\varphi_{\mu}^{ab}=\omega_{\mu}^{ab}\,,\qquad s\omega_{\mu}^{ab}=0\,,\label{brsgz}
\end{eqnarray}
where $(\bar{\varphi},\varphi)$ are a pair of complex commutating
fields, while $(\bar{\omega},\omega)$ are anti-commutating ones.
Now, the local version of the GZ action is then given by: 
\begin{eqnarray}
S_{\mathrm{GZ}}^{\mbox{\footnotesize{\it local}}} & = & S_{\mathrm{YM}}+\int d^{4}x\,\Bigl[\,\bar{\varphi}_{\mu}^{ac}\mathcal{M}^{ab}\varphi_{\mu}^{bc}-\bar{\omega}_{\mu}^{ac}\mathcal{M}^{ab}\omega_{\mu}^{bc}+\gamma^{2}gf^{abc}(\varphi_{\mu}^{ab}-\bar{\varphi}_{\mu}^{ab})A_{\mu}^{c}\Bigl]\,.\label{LocalGZ}
\end{eqnarray}
It is quite easy to notice that this term explicitly breaks the BRST
symmetry. Then, following the Zwanziger steps, in order to establish
a local, renormalizable and BRST-invariant theory, we define a most
general invariant action, which possesses action (\ref{LocalGZ})
as a particular physical case. Such desired action is then given by:
\begin{eqnarray}
S_{\mathrm{GZ}}^{\mbox{\footnotesize{\it local-inv}}} & = & S_{\mathrm{YM}}+s\int d^{4}x\,\Bigl[\,\bar{\omega}_{\mu}^{ac}\mathcal{M}^{ab}\varphi_{\mu}^{bc}+\bar{Q}_{\mu\nu}^{ab}\, D_{\mu}^{ac}\varphi_{\nu}^{cb}+{J}_{\mu\nu}^{ab}\, D_{\mu}^{ac}\bar{\omega}_{\nu}^{cb}+\bar{Q}_{\mu\nu}^{ab}J_{\mu\nu}^{ab}\Bigr]\nonumber \\
 & = & S_{\mathrm{YM}}+\int d^{4}x\,\Bigl\{\,\bar{\varphi}_{\mu}^{ac}\mathcal{M}^{ab}\varphi_{\mu}^{bc}-\bar{\omega}_{\mu}^{ac}\mathcal{M}^{ab}\omega_{\mu}^{bc}+f^{abc}(\partial_{\mu}\omega_{\nu}^{ae})(D_{\mu}^{bd}c^{d})\varphi_{\nu}^{ce}\nonumber \\
 &  & +\bar{J}_{\mu\nu}^{ab}\, D_{\mu}^{ac}\varphi_{\nu}^{cb}-\bar{Q}_{\mu\nu}^{ab}\left[D_{\mu}^{ac}\omega_{\nu}^{cb}+gf^{acd}(D_{\mu}^{bd}c^{d})\varphi_{\nu}^{cb}\right]+{J}_{\mu\nu}^{ab}\left[D_{\mu}^{ac}\bar{\varphi}_{\nu}^{cb}+gf^{acd}(D_{\mu}^{bd}c^{d})\bar{\omega}_{\nu}^{cb}\right]\nonumber \\
 &  & +\bar{Q}_{\mu\nu}^{ab}\, D_{\mu}^{ac}\bar{\omega}_{\nu}^{cb}+\left(\bar{J}_{\mu\nu}^{ab}{J}_{\mu\nu}^{ab}-\bar{Q}_{\mu\nu}^{ab}{Q}_{\mu\nu}^{ab}\right)\Bigl\}\,,\label{local-inv}
\end{eqnarray}
where, the set of external sources 
\begin{equation}
\{\bar{J}_{\mu\nu}^{ab},{J}_{\mu\nu}^{ab},\bar{Q}_{\mu\nu}^{ab},{Q}_{\mu\nu}^{ab}\}
\end{equation}
forms a BRST quartet structure, \textit{i.e.} 
\begin{eqnarray}
 &  & s{\bar{Q}}_{\mu\nu}^{ab}={\bar{J}}_{\mu\nu}^{ab}\,,\qquad s{\bar{J}}_{\mu\nu}^{ab}=0\,,\nonumber \\
 &  & sJ_{\mu\nu}^{ab}=Q_{\mu\nu}^{ab}\,,\qquad sQ_{\mu\nu}^{ab}=0\,,
\end{eqnarray}
being $(\bar{J},J)$ a pair of commutating sources and $(\bar{Q},Q)$
a pair of anti-commutating ones. The last term is a vacuum term permitted
by power-counting and it is necessary to obtain the gap equation (\ref{gapGZ})
by demanding that the vacuum energy, $\mathcal{E}$, is independent
of $\gamma^{2}$, \textit{i.e.} 
\begin{equation}
\frac{\partial\mathcal{E}}{\partial\gamma^{2}}=0\,.
\end{equation}
The original action (\ref{LocalGZ}) can be recovered from (\ref{local-inv})
when these external sources attain their \textit{physical values}.
Namely, 
\begin{equation}
\bar{J}_{\mu\nu}^{ab}\Bigl|_{phys}=-\bar{J}_{\mu\nu}^{ab}\Bigl|_{phys}=\gamma^{2}\,\delta^{ab}\delta_{\mu\nu}\,,\qquad\bar{Q}_{\mu\nu}^{ab}\Bigl|_{phys}=\bar{Q}_{\mu\nu}^{ab}\Bigl|_{phys}=0\,.
\end{equation}
After perform a linear shift on the $\omega_{\mu}^{ab}$ variable,
\begin{equation}
\omega_{\mu}^{ac}\to\omega_{\mu}^{ac}-(\mathcal{M}^{-1})^{ab}\left[\,\partial_{\nu}\left(gf^{bde}\varphi_{\mu}^{dc}\, D_{\nu}^{ef}c^{f}\right)-\gamma^{2}gf^{bdc}D_{\mu}^{de}c^{e}\,\right]\,,
\end{equation}
one can show that 
\begin{equation}
\left(S_{\mathrm{GZ}}^{\mbox{\footnotesize{\it local-inv}}}\right)\Bigr|_{phys}\equiv\, S_{\mathrm{GZ}}^{\mbox{\footnotesize{\it local}}}\,.
\end{equation}
It is important to emphasize here that the Zwanziger approach described
above has received some improvements in recent years. In \cite{Capri:2010hb,Capri:2011wp},
the model was formulated in such way that the BRST symmetry breaking
appears as a linear breaking, while in \cite{Dudal:2012sb}, the breaking
appears as a spontaneous breaking, instead of an explicit one%
\footnote{The renormalization of this formulation was already proven in \cite{CapriInProgress}.%
}.


\subsection{The $i$-particles}

As already seen, eq.~(\ref{i-propagator}) suggests that a theory
presenting Gribov-like propagators, can be rewritten in terms of $i$-particles,
\textit{i.e.} with propagations of ``particles'' with complex squared
masses. Now, let us take a look on this concept following the lines
outlined in \cite{Baulieu:2009ha}. Then, we will start with a scalar
field toy model exhibiting a confining Gribov-type propagator: 
\begin{equation}
S=\int d^{4}x\,\frac{1}{2}\,\Phi\left(-\partial^{2}+2\frac{\theta^{4}}{-\partial^{2}}\right)\Phi\,,
\end{equation}
where $\theta$ is a mass parameter playing an analogous role of the
Gribov parameter $\gamma$. The resulting propagator is the desired
Gribov-tipe: 
\begin{equation}
\langle\Phi(p)\Phi(-p)\rangle=\frac{p^{2}}{p^{4}+2\theta^{4}}\,,
\end{equation}
and it can be cast in a local form exactly like in the case of the
GZ model: 
\begin{equation}
S=\int d^{4}x\,\left[\frac{1}{2}\,\Phi(-\partial^{2})\Phi+\bar{\varphi}(-\partial^{2})\varphi+\theta^{2}\,\Phi(\varphi-\bar{\varphi})-\bar{\omega}(-\partial^{2})\omega\right]\,.
\end{equation}
As $(\bar{\varphi},\varphi)$ form a pair of complex field we can
decouple the real part from the theory. In fact, defining 
\begin{equation}
\varphi=\frac{1}{\sqrt{2}}(U+iV)\,,\qquad\bar{\varphi}=\frac{1}{\sqrt{2}}(U-iV)\,,
\end{equation}
one can write 
\begin{equation}
S=\int d^{4}x\,\left[\frac{1}{2}\,\Phi(-\partial^{2})\Phi+\frac{1}{2}\, V(-\partial^{2})V+\sqrt{2}i\theta^{2}\,\Phi V-\bar{\omega}(-\partial^{2})\omega+\frac{1}{2}\, U(-\partial^{2})U\right]\,.
\end{equation}
From now on, we will neglect the decoupled fields $(U,\bar{\omega},\omega)$
and we will diagonalize the action above by introducing the new field
variables: 
\begin{equation}
\Phi=\frac{1}{\sqrt{2}}(\lambda+\eta)\,,\qquad V=\frac{1}{\sqrt{2}}(\lambda-\eta)\,.
\end{equation}
Thus, we have 
\begin{equation}
S=\int d^{4}x\,\left[\frac{1}{2}\,\lambda(-\partial^{2}+i\sqrt{2}\theta^{2})\lambda+\frac{1}{2}\,\eta(-\partial^{2}-i\sqrt{2}\theta^{2})\eta\right]\,,
\end{equation}
and the propagators in terms of the new field variables stand by 
\begin{equation}
\langle\lambda(p)\lambda(-p)\rangle=\frac{1}{p^{2}+i\sqrt{2}\theta^{2}}\,,\qquad\langle\eta(p)\eta(-p)\rangle=\frac{1}{p^{2}-i\sqrt{2}\theta^{2}}\,.\label{i-action}
\end{equation}
From this expression one immediately sees that the fields $\lambda$
and $\eta$ correspond to the propagation of unphysical modes with
complex squared masses $\pm\sqrt{2}i\theta$. These are the so-called
$i$-particles of the model. Notice also that, despite the imaginary
terms, the action (\ref{i-action}) is Hermitian if we require that
$\lambda^{\dagger}=\eta$.\\
 \\
 We already argued that the excitations in terms of $i$-particles
are unphysical. Nevertheless, following \cite{Baulieu:2009ha}, physical
states can be introduced by constructing suitable composite operators
out of the fields $(\lambda,\eta)$ which exhibit desirable analyticity
properties, as encoded in the Källén-Lehmann spectral representation.
Such composite operators are obtained by requiring that the $i$-particles
fields enter in pairs, \textit{i.e.} the desired operator contains
as many fields of the type $\lambda$ as of the type $\eta$. It ensures
that in the corresponding correlation function only complex conjugate
pairs of $i$-particles propagate in the Feynman diagrams. In the
present case, the simplest example of a local composite operator with
the required physical properties is $\mathcal{O}(x)=\lambda(x)\eta(x)$.
In \cite{Baulieu:2009ha} it was shown that the correlation function
$\langle\mathcal{O}(p)\mathcal{O}(-p)\rangle$ has a well defined
Källén-Lehmann spectral representation%
\footnote{Here we present only the $4d$ result. The results for the spectral
density in $2d$ and $3d$ can also be found in \cite{Baulieu:2009ha}.%
}: 
\begin{equation}
\langle\mathcal{O}(p)\mathcal{O}(-p)\rangle=\int\frac{d^{4}q}{(2\pi)^{4}}\,\frac{1}{(p-q)^{2}-i\sqrt{2}\theta^{2}}\,\frac{1}{q^{2}+i\sqrt{2}\theta^{2}}\equiv\int_{\tau_{0}}^{\infty}d\tau\,\frac{\rho(\tau)}{\tau+p^{2}}\,,
\end{equation}
with 
\begin{equation}
\rho(\tau)=\frac{1}{(4\pi)^{2}}\,\sqrt{1-\frac{8\theta^{4}}{\tau^{2}}}\,,\qquad\tau_{0}=2\sqrt{2}\theta^{2}\,.
\end{equation}
Furthermore, a model with interacting $i$-particles is dealt in \cite{Capri:2012hh}.


\subsection{The replica model}

Let us now describe, in few words, the so-called replica model. It
was first introduced in \cite{Sorella:2010it} as an alternative way
to solve the Gribov problem. Here, we present the replica model by
starting with a scalar field with a quartic self-interaction term:
\begin{equation}
S=\int d^{4}x\,\left[\frac{1}{2}\,\Phi(-\partial^{2}+m^{2})\Phi+g\,\Phi^{4}\right]\,.
\end{equation}
Then, we define a replica of the action above, 
\begin{equation}
S'=\int d^{4}x\,\left[\frac{1}{2}\,\Phi'(-\partial^{2}+m^{2})\Phi'+g\,\Phi'^{4}\right]\,,
\end{equation}
with the same parameters $m$ and $g$. The two theories interact
by a soft term depending on a free parameter, say $\mu^{2}$, and
the replica model is then written as 
\begin{equation}
S_{replica}=\int d^{4}x\,\left[\frac{1}{2}\,\Phi(-\partial^{2}+m^{2})\Phi+\frac{1}{2}\,\Phi'(-\partial^{2}+m^{2})\Phi'+i\mu^{2}\,\Phi\Phi'+g\,\Phi^{4}+g\,\Phi'^{4}\right]\,.\label{replica}
\end{equation}
The replica model (\ref{replica}) enjoys a symmetry which guarantees
the existence of an unique mass parameter $m$ and an unique quartic
coupling constant $g$ for both sectors of the theory (the original
starting point model and its replica). This symmetry is often called
\textit{mirror symmetry} and its given by: 
\begin{equation}
\Phi\to\Phi'\,,\qquad\Phi'\to\Phi\,.
\end{equation}
Notice also that, when $m^{2}=0$, the propagator of $\Phi$ field
is of the Gribov type, with $\mu^{2}$ playing the role of the Gribov
parameter%
\footnote{For $m^{2}\neq0$ the propagator behaves like the most general RGZ
model \cite{Dudal:2008sp}.%
}. Thus, it can also be diagonalized in terms of $i$-particles. In
the deep ultraviolet regime, the two theories completely decouple
and we obtain that the $\Phi$ field is said deconfined exhibiting
a Yukawa-like propagator. The $\mu^{2}$ parameter might be fixed
by a gap equation like in the GZ model (\ref{gap}).


\section{The model: The $SU(3)$ Yang-Mills with dimension $2$ condensates}

We will begin with the pure Yang-Mills action in the Landau gauge,
eq.(\ref{YM}). Taking this action into account and trying not to
change the fundamental behavior in the ultraviolet regime we limit
ourselves to operators of ultraviolet dimension $2$. These type of
operators give rise to soft breaking terms. In particular, the mass
operator $\frac{1}{2}A_{\mu}^{a}A_{\mu}^{a}$ is well understood in
the context of the local composite operator method (LCO) \cite{Dudal:2003vv}
and is responsible for a Yukawa type propagator. We will focus into
another dimension $2$ operator. One that is only possible into $SU(N\geq3)$,
the operator constructed with the symmetric structure constant (see
Appendix B) and the gauge field $A_{\mu}^{a}$, \textit{i.e.} $\frac{1}{2}d^{abc}A_{\mu}^{b}A_{\mu}^{c}$.

It is perfectly possible to introduce these two operators in the quantum
action and study the renormalizability of both, which is done in Appendix
A. In order to introduce the operator $\frac{1}{2}d^{abc}A_{\mu}^{b}A_{\mu}^{c}$,
it is also necessary to introduce a BRST doublet of sources 
\begin{equation}
s\lambda^{a}=iJ^{a}\,,\qquad sJ^{a}=0\,.
\end{equation}
Then, the BRST invariant action including the dimension $2$ operator
$\frac{1}{2}d^{abc}A_{\mu}^{b}A_{\mu}^{c}$ is given by: 
\begin{eqnarray}
\Sigma & = & S_{\mathrm{YM}}+s\int d^{4}x\left(\frac{1}{2}\,\lambda^{a}d^{abc}A_{\mu}^{b}A_{\mu}^{c}-\frac{i\varepsilon}{2}\,\lambda^{a}J^{a}+\alpha\,\lambda^{a}\,\partial_{\mu}A_{\mu}^{a}\right)\nonumber \\
 & = & S_{\mathrm{YM}}+\int d^{4}x\left(\frac{i}{2}J^{a}d^{abc}A_{\mu}^{b}A_{\mu}^{c}+\lambda^{a}d^{abc}(\partial_{\mu}c^{b})A_{\mu}^{c}-\frac{g}{2}f^{abc}d^{cde}\,\lambda^{a}c^{b}A_{\mu}^{d}A_{\mu}^{e}+\frac{\varepsilon}{2}J^{a}J^{a}\right)\nonumber \\
 &  & \phantom{S_{\mathrm{YM}}}+\,\alpha\int d^{4}x\left(iJ^{a}\partial_{\mu}A_{\mu}^{a}+\lambda^{a}\partial_{\mu}D_{\mu}^{ab}c^{b}\right)\,.\label{simetrica}
\end{eqnarray}
The $\alpha$ terms are necessary by algebraic renormalization consistency,
due to the fact that there is any symmetry to exclude such terms%
\footnote{Also, in Landau gauge these extra terms can be absorbed by performing
a linear shift of the fields variables $b^{a}$ and

\begin{center}
$\bar{c}^{a}$:$b{a}\to b{a}+i\alpha J{a},\bar{c}{a}\to\bar{c}{a}+i\alpha\lambda{a}.$ 
\par\end{center}%
}. At this point, it is important to emphasize that, although we introduce
the operator consistent with the study of renormalizability via a
BRST doublet of sources, here we are interested in bringing the sources
for your physical value, following the same procedure of Gribov-Zwanziger
\cite{Zwanziger:1989mf,Zwanziger:1992qr}. A possible condensation
of this operator will give rise to a non-zero expectation value in
the vacuum of $J^{a}$%
\footnote{We are not calculating the value of this condensate. We limit ourselves
in this work to study the consequences of its existence.%
}. It is necessary a non-zero expected value for $J^{a}$ so as to
satisfy the Gribov condition (\ref{definicao}), which we will show
in the next section. Another possible way to obtain this value is
by the LCO method \cite{LCOLemes1,LCOLemes2,LCOLemes3} which uses
the technique of effective potential. We emphasize that it is not
necessary to implement the LCO here due to the Gribov condition.

Thus, it implies that we have to choose a direction for $J^{a}$.
To guide implementation of this choice let analyze some properties
of the group SU(3) (Appendix \ref{su3}). We can see that the gauge
field $A_{\mu}$ can be expanded in a base of the $SU(3)$ generators
as 
\begin{equation}
A_{\mu}=\sum_{a=1}^{8}A_{\mu}^{a}\frac{\lambda^{a}}{2}=\sum_{a\neq3,8}A_{\mu}^{a}\frac{\lambda^{a}}{2}+A_{\mu}^{+}\,\lambda_{+}+A_{\mu}^{-}\,\lambda_{-}\,,
\end{equation}
where 
\begin{equation}
A_{\mu}^{\pm}=\frac{1}{2}\left(\frac{1}{\sqrt{3}}\, A_{\mu}^{8}\pm A_{\mu}^{3}\right)\,.
\end{equation}
And the pure $A$-field sector of the $SU(3)$ Yang-Mills action,
\textit{i.e.} 
\begin{equation}
S_{A\mbox{\footnotesize-field}}=\frac{1}{4}\int d^{4}x\, F_{\mu\nu}^{a}F_{\mu\nu}^{a}\,,
\end{equation}
is left invariant by the transformations%
\footnote{Of course that the symmetry can be extended by the ghost and gauge
fixing sectors of action (\ref{YM})%
} 
\begin{eqnarray}
(A_{\mu}^{1},A_{\mu}^{2}) & \to & (A_{\mu}^{1},A_{\mu}^{2})\,,\nonumber \\
(A_{\mu}^{+},A_{\mu}^{4},A_{\mu}^{5}) & \to & (-A_{\mu}^{-},A_{\mu}^{7},A_{\mu}^{6})\,,\nonumber \\
(A_{\mu}^{-},A_{\mu}^{6},A_{\mu}^{7}) & \to & (-A_{\mu}^{+},-A_{\mu}^{5},-A_{\mu}^{4})\,.\label{symm}
\end{eqnarray}
Notice that these transformations interchange the groups $SU(2)_{\mathrm{II}}$
and $SU(2)_{\mathrm{III}}$ and it reminds us the mirror symmetry
of the replica model described in the preceding section. However,
we can not identify these two subsectors of the theory as a replica
model, at least not as originally conceived in \cite{Sorella:2010it},
because of the presence of hard interaction terms. Furthermore, the
only direction that maintains the symmetry (\ref{symm}) is the direction
$3$. Then, choosing

\begin{equation}
\langle J^{a}\rangle=m^{2}\delta^{a3}\,,\label{eq:sourcephyvalue}
\end{equation}
we have: 
\begin{equation}
i\langle J^{a}\rangle\, d^{abc}A_{\mu}^{b}A_{\mu}^{c}=\frac{im^{2}}{3}\,(A_{\mu}^{+}A_{\mu}^{+}-A_{\mu}^{-}A_{\mu}^{-})+\frac{im^{2}}{2}\,(A_{\mu}^{4}A_{\mu}^{4}-A_{\mu}^{7}A_{\mu}^{7})+\frac{im^{2}}{2}\,(A_{\mu}^{5}A_{\mu}^{5}-A_{\mu}^{6}A_{\mu}^{6})\,.
\end{equation}

And, with choice (\ref{eq:sourcephyvalue}), the action (\ref{simetrica})
gives rise to a propagator of the form: 
\begin{eqnarray}
\langle A_{\mu}^{a}(k)A_{\nu}^{b}(-k)\rangle & = & \biggl[\frac{1}{k^{2}}\sum_{i=1}^{2}\delta^{ai}\delta^{bi}+\frac{1}{k^{2}+i\frac{m^{2}}{2}}\sum_{i=4}^{5}\delta^{ai}\delta^{bi}+\frac{1}{k^{2}-i\frac{m^{2}}{2}}\sum_{i=6}^{7}\delta^{ai}\delta^{bi}\nonumber \\
 &  & +\frac{k^{2}}{k^{4}+\frac{m^{4}}{3}}(\delta^{a8}\delta^{b8}+\delta^{a3}\delta^{b3})-i\frac{m^{2}}{\sqrt{3}}\frac{1}{k^{4}+\frac{m^{4}}{3}}(\delta^{a8}\delta^{b3}+\delta^{a3}\delta^{b8})\biggr]\,\theta_{\mu\nu}(k)\,,\label{pro-gribov}
\end{eqnarray}
with $\theta_{\mu\nu}(k)$ being the transverse projector, 
\begin{equation}
\theta_{\mu\nu}(k)=\left(\delta_{\mu\nu}-\frac{k_{\mu}k_{\nu}}{k^{2}}\right)\,.
\end{equation}
This calculation is done in the Gell-Mann representation \cite{Narison:2002pw},
see Appendix B. Although (\ref{eq:sourcephyvalue}) explicitly breaking
the group, this is a soft breaking and in the ultraviolet regime is
expected that the group structure is recovered. Note that we get a
propagator which has the \textit{i-particle} structure (\ref{i-propagator}).

It is important to stress here that in spite of have two massless
poles in the propagator it is not necessary that we have physical
particles directly associated to this propagator. The interaction
terms will mix the \textit{i-particles} with the massless ones. This
property can be easily observed if we rewrite the propagator as: 
\begin{eqnarray}
\langle A_{\mu}^{a}(k)A_{\nu}^{b}(-k)\rangle & = & \biggl\{\frac{1}{k^{4}+\frac{m^{4}}{4}}(k^{2}\delta^{ab}-im^{2}d^{ab3})+\frac{m^{4}}{4(k^{4}+\frac{m^{4}}{4})}\biggl[\frac{1}{k^{2}}\sum_{i=1}^{2}\delta^{ai}\delta^{bi}\nonumber \\
 &  & -\frac{1}{3}\,\frac{k^{2}}{k^{4}+\frac{m^{4}}{3}}\,(\delta^{a8}\delta^{b8}+\delta^{a3}\delta^{b3})+i\frac{m^{2}}{3\sqrt{3}}\frac{1}{k^{4}+\frac{m^{4}}{3}}\,(\delta^{a8}\delta^{b3}+\delta^{a3}\delta^{b8})\biggr]\biggr\}\,\theta_{\mu\nu}(k)\,,\label{pro-gribov2}
\end{eqnarray}
where $d^{ab3}$ are symmetric structure constants in the $3$ direction.
In particular, for further use, we also display: 
\begin{eqnarray}
\langle A_{\mu}^{1}(k)A_{\nu}^{1}(-k)\rangle & = & \langle A_{\mu}^{2}(k)A_{\nu}^{2}(-k)\rangle=\frac{1}{k^{2}}\,\theta_{\mu\nu}(k)\,,\nonumber \\
\langle A_{\mu}^{4}(k)A_{\nu}^{4}(-k)\rangle & = & \langle A_{\mu}^{5}(k)A_{\nu}^{5}(-k)\rangle=\frac{1}{k^{2}+i\frac{m^{2}}{2}}\,\theta_{\mu\nu}(k)\,,\nonumber \\
\langle A_{\mu}^{6}(k)A_{\nu}^{6}(-k)\rangle & = & \langle A_{\mu}^{7}(k)A_{\nu}^{7}(-k)\rangle=\frac{1}{k^{2}-i\frac{m^{2}}{2}}\,\theta_{\mu\nu}(k)\,,\nonumber \\
\langle A_{\mu}^{3}(k)A_{\nu}^{3}(-k)\rangle & = & \langle A_{\mu}^{8}(k)A_{\nu}^{8}(-k)\rangle=\frac{k^{2}}{k^{4}+\frac{m^{4}}{3}}\,\theta_{\mu\nu}(k)\,,\nonumber \\
\langle A_{\mu}^{3}(k)A_{\nu}^{8}(-k)\rangle & = & \langle A_{\mu}^{8}(k)A_{\nu}^{3}(-k)\rangle=-i\frac{m^{2}}{\sqrt{3}}\frac{1}{k^{4}+\frac{m^{4}}{3}}\,\theta_{\mu\nu}(k)\,.
\end{eqnarray}
At this point some considerations about the explicitly breaking of
the $SU(3)$ are necessary. First, as is shown in Appendix \ref{su3},
the Gell-Mann matrices grouped as in (\ref{3xSU(3)}) have the same
algebraic properties as the Pauli matrices and so determine three
natural $SU(2)$ subalgebras. So taking 
\begin{eqnarray*}
h_{1} & = & \lambda^{3},\hspace{0.2cm}h_{2}=\lambda^{8},\hspace{0.2cm}e_{\pm}^{1}=\lambda^{1}\pm i\lambda^{2},\hspace{0.2cm}e_{\pm}^{2}=\lambda^{6}\pm i\lambda^{7},\hspace{0.2cm}e_{\pm}^{3}=\lambda^{4}\pm i\lambda^{5},
\end{eqnarray*}
it is easy to observe that the $e_{\pm}^{i}$ obeys the following
algebra 
\begin{eqnarray}
\left[h_{1},h_{2}\right] & = & 0,\hspace{0.2cm}\left[\sqrt{3}h_{2}+h_{1},e_{\pm}^{1}\right]=\pm e_{\pm}^{1},\hspace{0.2cm}\left[\sqrt{3}h_{2}+h_{1},e_{\pm}^{2}\right]=\mp\frac{1}{2}e_{\pm}^{2},\hspace{0.2cm}\left[\sqrt{3}h_{2}+h_{1},e_{\pm}^{3}\right]=\pm\frac{1}{2}e_{\pm}^{3},\hspace{0.2cm}\nonumber \\
\left[\sqrt{3}h_{2}-h_{1},e_{\pm}^{1}\right] & = & \mp e_{\pm}^{1},\hspace{0.2cm}\left[\sqrt{3}h_{2}-h_{1},e_{\pm}^{2}\right]=\pm\frac{1}{2}e_{\pm}^{2},\hspace{0.2cm}\left[\sqrt{3}h_{2}-h_{1},e_{\pm}^{3}\right]=\mp\frac{1}{2}e_{\pm}^{3},\hspace{0.2cm}\nonumber \\
\left[e_{+}^{1},e_{-}^{1}\right] & = & 2h_{1},\hspace{0.2cm}\left[e_{+}^{2},e_{-}^{2}\right]=\sqrt{3}h_{2}-h_{1},\hspace{0.2cm}\left[e_{+}^{3},e_{-}^{3}\right]=\sqrt{3}h_{2}+h_{1}.\label{algebra3su2}
\end{eqnarray}
This give rise to the well know weight diagram \cite{Narison:2002pw}.
Moreover, from the properties of $SU(2)$ representations we know
that $2p=m_{1};\sqrt{3}q-p=m_{2};\sqrt{3}q+p=m_{3}$, where $(p,q)$
corresponds respectively to the eigenvalues of $h_{1}$ and $h_{2}$
ordered as points in $\mathbb{R}^{2}$. Assuming that the trace of
the propagator must be real, the natural choice of the direction of
the breaking in $h_{1}$ give rise to two goldstone bosons associated
to $e_{\pm}^{1}$. The remaining 2 sets of $SU(2)$ corresponds to
a similar structure as the replica model \cite{Sorella:2010it}. It
is important to stress here that this is not a model for confinement
in $SU(3)$. This is essentially an alternative mechanism for Gribov
that presents confinement in the remaining two $SU(2)$ groups and
has some defined observables associated to the remaining $SU(2)$
groups. In section \ref{sec:observables} we will return to these
issues in order to define the relationship between observables and
the remaining group structure.

In order to make clear the importance of each operator and their relation
to the two types of solutions came from the Schwinger-Dyson equations,
the scaling type and the decoupling one, in next section we discuss
the relation between these solutions and the operators $\frac{1}{2}d^{abc}A_{\mu}^{b}A_{\mu}^{c}$
and $\frac{1}{2}A_{\mu}^{a}A_{\mu}^{a}$.


\section{Taking into account the Gribov copies. The scaling type solution
for the gluon and ghost propagators}

\label{sec:gapequation}

In order to offer a better understanding of the model, a dynamical
framework for the parameter $\langle J^{a}\rangle=m^{2}\delta^{a3}$
should be provided, \textit{i.e.} we should be able to establish a
gap equation for that parameter, allowing us to express $m^{2}$ as
a function of the coupling constant g, as in the Gribov approximation
\cite{Gribov:1977wm} or in the GZ theory \cite{Zwanziger:1991ac,Zwanziger:1992qr}.
The most immediate way to achieve a meaningful gap equation for $m^{2}$
is following the steps detailed in (\ref{definicao})$\rightarrow$(\ref{gap}),
which consists in restricting the domain of integration in the functional
integral to the Gribov region with no-pole condiction (\ref{sigma}).
Note that this condiction relies on the observation that the Faddeev-Popov
operators are invertible in Gribov region and their inverse are nothing
but the twopoint ghost functions (\ref{eq:twopointghosparam}).

Therefore, in our case, we have to calculate the twopoint ghost functions
(\ref{eq:twopointghosparam}), with the propagator (\ref{pro-gribov}).
And the no-pole condition is implemented (see \cite{Sorella:2010it}
too) by stating that 
\begin{equation}
\sigma(0,A)=1,\label{gap-replica}
\end{equation}
which yields the gap equation determining the parameter $m^{2}$,
or equivalently $\langle J^{a}\rangle\langle J^{a}\rangle$. After
some calculation the gap equation yields: 
\begin{equation}
1=g^{2}\frac{3}{4}\int\frac{d^{D}k}{(2\pi)^{D}}\frac{1}{k^{4}+\frac{m^{4}}{4}}+g^{2}\frac{1}{16}\int\frac{d^{D}k}{(2\pi)^{D}}\frac{m^{4}}{k^{2}}(\frac{1}{k^{4}+\frac{m^{4}}{4}})(\frac{1}{k^{4}+\frac{m^{4}}{3}})\;,\label{gapp}
\end{equation}
where dimensional regularization, $D=4-\epsilon$, has been employed,
and we make use of $d^{aac}\langle J^{c}\rangle=0$%
\footnote{This general property can be easily seen, for example, in the Gell-Mann
representation for the generators.%
}.

The gap equation (\ref{gapp}) enables us to express the parameter
$m^{4}$ as a function of the coupling constant $g$. It is clear
that the second integral is absolutely convergent and do not change
the fact that $m$ is determined as a function of a regularization
mass $\Lambda$.

Let us now see that our model also recovers the ghost propagator which
is enhanced in the infrared as in the usual Gribov approach that is
made in detail in the review \cite{Sobreiro:2005ec}. The ghost propagator
is given by (\ref{eq:twopointghosparam}) with $\sigma(p^{2},A)$
defined by (\ref{defregion}). As our gauge field propagator is (\ref{pro-gribov}),
we have (with $N=3$):

\[
\sigma(p^{2},m^{2})=3g^{2}\frac{p_{\mu}p_{\nu}}{p^{2}}\int\frac{d^{D}k}{(2\pi)^{D}}\frac{1}{(p-k)^{2}}\left[\frac{k^{2}}{k^{4}+\frac{m^{4}}{4}}+(\frac{m^{4}}{12(k^{4}+\frac{m^{4}}{4})(k^{4}+\frac{m^{4}}{3})})\right]\,\theta_{\mu\nu}(k)\;,
\]

Let us analyze the infrared behavior, $k\approx0$, of $(1-\sigma(p^{2},m^{2}))$.
Making use of the gap equation (\ref{gapp}), which can be rewritten
from Lorentz covariance as: 
\begin{equation}
1=3g^{2}\frac{p_{\mu}p_{\nu}}{p^{2}}\int\frac{d^{D}k}{(2\pi)^{D}}\left[\frac{1}{k^{4}+\frac{m^{4}}{4}}+(\frac{m^{4}}{12k^{2}(k^{4}+\frac{m^{4}}{4})(k^{4}+\frac{m^{4}}{3})})\right]\,\theta_{\mu\nu}\;,\label{gapp-1}
\end{equation}

We obtain for $(1-\sigma(p^{2},m^{2}))$:

\begin{align}
(1-\sigma(p^{2},m^{2})) & =3g^{2}\frac{p_{\mu}p_{\nu}}{p^{2}}\int\frac{d^{D}k}{(2\pi)^{D}}\left(1-\frac{k^{2}}{(p-k)^{2}}\right)\left[\frac{1}{k^{4}+\frac{m^{4}}{4}}+(\frac{m^{4}}{12k^{2}(k^{4}+\frac{m^{4}}{4})(k^{4}+\frac{m^{4}}{3})})\right]\,\theta_{\mu\nu}(k)\;\nonumber \\
 & =3g^{2}\frac{p_{\mu}p_{\nu}}{p^{2}}\int\frac{d^{D}k}{(2\pi)^{D}}\left(\frac{p^{2}-2pk}{(p-k)^{2}}\right)\left[\frac{1}{k^{4}+\frac{m^{4}}{4}}+(\frac{m^{4}}{12k^{2}(k^{4}+\frac{m^{4}}{4})(k^{4}+\frac{m^{4}}{3})})\right]\,\theta_{\mu\nu}(k)\;\\
 & =3g^{2}\frac{p_{\mu}p_{\nu}}{p^{2}}\mathcal{P}_{\mu\nu}(p)\;,\nonumber 
\end{align}

where

\begin{equation}
\mathcal{P}_{\mu\nu}(p)=\int\frac{d^{D}k}{(2\pi)^{D}}\left(\frac{p^{2}-2pk}{(p-k)^{2}}\right)\left[\frac{1}{k^{4}+\frac{m^{4}}{4}}+(\frac{m^{4}}{12k^{2}(k^{4}+\frac{m^{4}}{4})(k^{4}+\frac{m^{4}}{3})})\right]\,\theta_{\mu\nu}(k)\;.
\end{equation}

From this expression, one sees that $\mathcal{P}_{\mu\nu}(p)$ is
convergent and non singular at $p=0$. It follows that, for $p\approx0$:

\begin{equation}
\mathcal{P}_{\mu\nu}(p)_{p\rightarrow0}\approx p^{2}\int\frac{d^{D}k}{(2\pi)^{D}}\frac{1}{k{}^{2}}\left[\frac{1}{k^{4}+\frac{m^{4}}{4}}+(\frac{m^{4}}{12k^{2}(k^{4}+\frac{m^{4}}{4})(k^{4}+\frac{m^{4}}{3})})\right]\,\theta_{\mu\nu}(k)\;.
\end{equation}

Thus, it follows that, for small values of the momentum 
\begin{equation}
(1-\sigma(p^{2},m^{4}))\Big|_{k^{2}\approx0}={\cal C}p^{2}\;,\label{kzero}
\end{equation}
and we recover the ghost propagator which is enhanced in the infrared
$\langle\overline{c}^{a}(p)c^{b}(-p)\rangle\approx\frac{1}{p^{4}}$
(see equation (\ref{eq:twopointghosparam})). We have thus recovered
the so-called scaling solution, \textit{i.e.} a suppressed gluon propagator
which vanishes at the origin (\ref{pro-gribov}), and enhanced ghosts,
which corresponds to the solution of the GZ theory.

\subsection{Taking into account $\frac{1}{2}A_{\mu}^{a}A_{\mu}^{a}$ and the
decoupling type solution.}

Recent lattice numerical simulations \cite{Cucchieri:2007rg,Cucchieri:2008fc,Cucchieri:2008mv,Cucchieri:2009zt,Bogolubsky:2009dc,Bogolubsky:2009qb,Dudal:2010tf}
indicates a gluon propagator which is suppressed in the infrared and
which attains a finite non-vanishing value at zero momentum, while
the ghost propagator turns out to be not enhanced, i.e. $<\overline{c}^{a}(k)c^{b}(-k)>\approx\frac{1}{k^{2}}$.
This behaviour is know as the decoupling solution and has also been
obtained from the analysis of the Schwinger-Dyson equations \cite{Aguilar:2004sw,Aguilar:2008xm,Boucaud:2008ky}.
This solution appears in the Gribov-Zwanziger theory when the dynamics
of the lacalizing fields is taken into account. In our model this
behaviour is associated to the mass operator $\frac{1}{2}A_{\mu}^{a}A_{\mu}^{a}$.
Before presenting the equations that characterize the stability of
this operator it is important to emphasize that the tad pole presented
in (\ref{separa2}) despite being ultraviolet convergent has problems
in infrared, but not necessarily at $k=0$. The best chance to solve
this divergency is introducing a mass term in the action.

The action $\Sigma$ with a mass term is given by: 
\begin{equation}
\Sigma_{\mu}=\Sigma+\frac{\mu^{2}}{2}\int{d^{4}x}\, A_{\mu}^{a}A_{\mu}^{a}\,,\label{mass}
\end{equation}
where $\Sigma$ was defined in (\ref{simetrica}). The BRST variation
of the mass term, turns out to be proportional to the equation of
motion of $b^{a}$ i.e 
\begin{equation}
\mathcal{S}(\Sigma_{\mu})=0-\mu^{2}\int{d^{4}x}\,(\partial_{\mu}c^{a})A_{\mu}^{a}=-i\mu^{2}\int{d^{4}x}\, c^{a}\frac{\delta\Sigma_{m}}{\delta b^{a}}\,,\label{mass-sim}
\end{equation}
modifying the Slavnov-Taylor to 
\begin{eqnarray}
{\overline{{\cal {S}}}}(\Sigma_{\mu}) & = & \int d^{4}x\lbrace\frac{\delta\Sigma_{\mu}}{\delta\Omega_{\mu}^{a}}\frac{\delta\Sigma_{\mu}}{\delta A_{\mu}^{a}}+\frac{\delta\Sigma_{\mu}}{\delta L^{a}}\frac{\delta\Sigma_{\mu}}{\delta c^{a}}+ib^{a}\frac{\delta\Sigma_{\mu}}{\delta\overline{c}^{a}}+ij^{a}\frac{\delta\Sigma_{\mu}}{\delta\lambda^{a}}+i\mu^{2}c^{a}\frac{\delta\Sigma_{\mu}}{\delta b^{a}}\rbrace\nonumber \\
 & = & 0.
\end{eqnarray}
It is now clear that the gauge propagator changes to a propagator
in close relation to the one obtained from the refined Gribov-Zwanziger
theory \cite{Dudal:2007cw,Dudal:2008sp,Dudal:2008rm}. The new propagator
is given by: 
\begin{eqnarray}
<A_{\mu}^{a}(k)A_{\nu}^{b}(-k)> & = & [(\frac{1}{(k^{2}+\mu^{2})^{2}+\frac{m^{4}}{4}})((k^{2}+\mu^{2})\delta^{ab}-im^{2}d^{ab3})\nonumber \\
 & + & \frac{m^{4}}{4((k^{2}+\mu^{2})^{2}+\frac{m^{4}}{4})}\{(\frac{1}{k^{2}+\mu^{2}})\sum_{i=1}^{2}\delta^{ai}\delta^{bi}-\frac{1}{3}(\frac{k^{2}+\mu^{2}}{(k^{2}+\mu^{2})^{2}+\frac{m^{4}}{3}})(\delta^{a8}\delta^{b8}+\delta^{a3}\delta^{b3})\nonumber \\
 & + & i\frac{m^{2}}{\sqrt{3}}\frac{1}{3}(\frac{1}{(k^{2}+\mu^{2})^{2}+\frac{m^{4}}{3}})(\delta^{a8}\delta^{b3}+\delta^{a3}\delta^{b8})\}]\theta_{\mu\nu},
\end{eqnarray}
where $m^{4}$ and $\mu^{2}$ are obtained by Gribov conditions presented
in Appendix A3. This expression has a finite nonvanishing value at
zero momentum characterizing the decoupling type solution. It is important
to emphasize here that lattice simulation results that have obtained
the behaviour of the propagator are making use of the trace of the
propagator. Remembering that $d^{aab}=0$, the dominating term that
lattice is capable to see is $\frac{k^{2}+\mu^{2}}{k^{4}+2\mu^{2}k^{2}+\frac{m^{4}}{4}+\mu^{4}}$.
It is clear that the dynamical origem of this mass parameter needs
more explanation and we pretend to do this into a future work.

\section{Local composite operator and the Källén-Lehmann spectral representation}

\label{sec:observables}

Remove the gauge fields of the physical spectrum of the theory is
not enough for a model that attempts to offer an alternative, at least
in part, to the original Gribov question. It is also necessary to
present a candidate for physical observable that displays the Källén-Lehmann
spectral representation and corresponds to an invariant composite
operator. It is pointed out in \cite{Baulieu:2009ha,Sorella:2010it}
that a local composite operator constructed with {\textit{i}}-particles
\cite{Baulieu:2009ha} has one-loop correlation function that exibits
the Källén-Lehmann spectral representation, i.e. 
\begin{equation}
{\cal {I}}(p^{2})=\int{\frac{d^{4}k}{(2\pi)^{4}}}\frac{1}{((p-k)^{2}+i\sqrt{2}\nu^{2})(k^{2}-i\sqrt{2}\nu^{2})},\label{int-observ}
\end{equation}
exhibits a spectral representation, as: 
\begin{equation}
{\cal {I}}(p^{2})-{\cal {I}}(0)=\int_{2\sqrt{2}\nu^{2}}^{\infty}d\upsilon\rho(\upsilon)\{\frac{1}{\upsilon+p^{2}}-\frac{1}{\upsilon}\},
\end{equation}
where the spectral density $\rho(\upsilon)=\frac{1}{16\pi^{2}}\frac{\sqrt{\upsilon^{2}-8\nu^{4}}}{\upsilon}$
is positive in the range of integration. This property help us to
find an operator that has the desired analyticity properties. Now
let us return to the breaking of the $SU(3)$ group in order to understand
the mechanism that permits the existence of physical observables.
First of all, one of the desired aspects for one observable is not
only that has Källén-Lehmann spectral representation of a particle
but also does not carries color index and be gauge invariant. It is
clear that the last requirement is not possible due to the explicitly
breaking of the BRST symmetry. This is also a problem in the original
Gribov-Zwanziger due to the same problem. i.e. the explicitly symmetry
breaking. In our case another question emerges, the explicitly breaking
of the group structure. Fortunately the last question will be the
answer in order to obtain observables that does not carry color index.
At least of the remaining group structure.

Analyzing the {\textit{i}}-particles concept presented in \cite{Baulieu:2009ha,Sorella:2010it}
we can see that it is impossible to obtain from a Gribov type propagator
an observable that do not corresponds to an integral of the type (\ref{int-observ}).
This does not only suggests that the observables must be constructed
taking into account {\textit{i}}-particle type correlators in order
to obtain a real particle pole, but also indicates that a mechanism
in order to do that must mix two different types of particles as done
in \cite{Sorella:2010it}. The natural candidates that emerges as
possible observables can be associated to the remaining 2 sets of
$\mathcal{L}(SU(2))$, in particular it is convenient to define the
quantities $E_{\mu\nu}^{3+}\equiv\frac{1}{\sqrt{2}}(F_{\mu\nu}^{4}+iF_{\mu\nu}^{5})$,
$E_{\mu\nu}^{3-}\equiv\frac{1}{\sqrt{2}}(F_{\mu\nu}^{4}-iF_{\mu\nu}^{5})$,
$E_{\mu\nu}^{2+}\equiv\frac{1}{\sqrt{2}}(F_{\mu\nu}^{6}+iF_{\mu\nu}^{7})$
and $E_{\mu\nu}^{2-}\equiv\frac{1}{\sqrt{2}}(F_{\mu\nu}^{6}-iF_{\mu\nu}^{7})$
that obeys the same algebra as presented in (\ref{algebra3su2}).
From these operators it is convenient to define one possible candidate
to observable as: 
\begin{eqnarray}
\phi & \equiv & E_{\mu\nu}^{2+}E_{\mu\nu}^{3+}+E_{\mu\nu}^{3-}E_{\mu\nu}^{2-}\nonumber \\
\phi & = & F_{\mu\nu}^{4}F_{\mu\nu}^{6}-F_{\mu\nu}^{5}F_{\mu\nu}^{7}.\label{observ1}
\end{eqnarray}
It should be noted that in spite of the breaking of the BRST symmetry,
the candidate to observable must be a singlet. It is important to
remember that the $E_{\mu\nu}$ essentially obeys the same group properties
as defined in (\ref{algebra3su2}). Remenbering that we are performin
the breaking in diretion $3$ or in the notation presented in (\ref{algebra3su2})
$h_{1}$, the algebra between $h_{1}$ and $e_{\pm}^{2}$,$e_{\pm}^{2}$
is given by: 
\begin{equation}
[h_{1},e_{\pm}^{2}]=\mp\frac{1}{2}e_{\pm}^{2}\hspace{1cm}[h_{1},e_{\pm}^{3}]=\pm\frac{1}{2}e_{\pm}^{3},\label{algebra-e+-}
\end{equation}
which makes the result algebraic $[h_{1},e_{+}^{2}e_{+}^{3}+e_{-}^{3}e_{-}^{2}]=0$
easier to obtain and proving that this operator is invariant under
the remaining group charge defined by the algebra (\ref{algebra-e+-}).
Taking into account the propagators as: 
\begin{eqnarray}
<A_{\mu}^{1}(k)A_{\nu}^{1}(-k)> & = & <A_{\mu}^{2}(k)A_{\nu}^{2}(-k)>=\frac{1}{k^{2}}\theta_{\mu\nu}(k)\nonumber \\
<A_{\mu}^{4}(k)A_{\nu}^{4}(-k)> & = & <A_{\mu}^{5}(k)A_{\nu}^{5}(-k)>=\frac{1}{k^{2}+i\frac{m^{2}}{2}}\theta_{\mu\nu}(k)\nonumber \\
<A_{\mu}^{6}(k)A_{\nu}^{6}(-k)> & = & <A_{\mu}^{7}(k)A_{\nu}^{7}(-k)>=\frac{1}{k^{2}-i\frac{m^{2}}{2}}\theta_{\mu\nu}(k)\nonumber \\
<A_{\mu}^{3}(k)A_{\nu}^{3}(-k)> & = & <A_{\mu}^{8}(k)A_{\nu}^{8}(-k)>=\frac{k^{2}}{k^{4}+\frac{m^{4}}{3}}\theta_{\mu\nu}(k)\nonumber \\
<A_{\mu}^{3}(k)A_{\nu}^{8}(-k)> & = & <A_{\mu}^{8}(k)A_{\nu}^{3}(-k)>=-i\frac{m^{2}}{\sqrt{3}}\frac{1}{k^{4}+\frac{m^{4}}{3}}\theta_{\mu\nu}(k),
\end{eqnarray}
the first candidate which has the desired one-loop correlation function
\cite{Baulieu:2009ha} is given by: 
\begin{eqnarray}
<\phi(k)\phi(-k)> & = & 12\int_{m^{2}}^{\infty}d\upsilon\frac{\rho(\upsilon)}{\upsilon+k^{2}},\nonumber \\
\rho(\upsilon) & = & \frac{\sqrt{\upsilon^{2}-m^{4}}(\upsilon^{2}+m^{4})}{\pi^{2}\upsilon}.
\end{eqnarray}

There is another observable with different value for the mass of the
condensate. In order to understand the second candidate to an observable,
i.e. that presents the Källén-Lehmann spectral representation of a
particle, it is convenient to consider the quadratic part of the action,
in particular the term with $A_{\mu}^{3}$ and $A_{\mu}^{8}$. 
\begin{equation}
S_{3,8}=\int d^{4}x\{\frac{1}{2}A_{\mu}^{3}(-\partial^{2})A_{\mu}^{3}+\frac{1}{2}A_{\mu}^{8}(-\partial^{2})A_{\mu}^{8}+m^{2}\frac{i}{\sqrt{3}}A_{\mu}^{3}A_{\mu}^{8}\},\label{38}
\end{equation}
where we have already taken into account the Landau gauge conditions,$\partial_{\mu}A_{\mu}^{3}=0$
and $\partial_{\mu}A_{\mu}^{8}=0$. This sector of the action can
be diagonalized trivialy making the field redefinition: 
\begin{eqnarray}
U_{\mu} & = & \frac{1}{\sqrt{2}}(A_{\mu}^{3}+A_{\mu}^{8})\nonumber \\
V_{\mu} & = & \frac{1}{\sqrt{2}}(-A_{\mu}^{3}+A_{\mu}^{8}).
\end{eqnarray}
Therefore 
\begin{equation}
S_{3,8}=\int d^{4}x\{\frac{1}{2}U_{\mu}(-\partial^{2}+i\frac{m^{2}}{\sqrt{3}})U_{\mu}+\frac{1}{2}V_{\mu}(-\partial^{2}-i\frac{m^{2}}{\sqrt{3}})V_{\mu}\},
\end{equation}
which describes again the \textit{i-particle} structure. In order
to write a physical operator that has the desired structure, it is
sufficient to study the operator 
\begin{equation}
\chi=(\partial_{\mu}U_{\nu}-\partial_{\nu}U_{\mu})(\partial_{\mu}V_{\nu}-\partial_{\nu}V_{\mu}),
\end{equation}
where its two-point correlation function can be cast in the form of
a Källén-Lehmann spectral representation of a physical particle, 
\begin{eqnarray}
<\chi(k)\chi(-k)> & = & \frac{3}{8}\int_{2\frac{m^{2}}{\sqrt{3}}}^{\infty}d\upsilon\frac{\rho(\upsilon)}{\upsilon+k^{2}},\nonumber \\
\rho(\upsilon) & = & \frac{\sqrt{\upsilon^{2}-\frac{4}{3}m^{4}}(\upsilon^{2}+\frac{4}{3}m^{4})}{\pi^{2}\upsilon}.
\end{eqnarray}
Again it is convenient to observe that this operator can be associated
to the abelian subgroup of the original $SU(3)$ group and it is clear
that the mass pole only depends on the mass gap of the \textit{i-particle}.
Moreover it is important to stress that due to the fact that the insertion
is a soft broken, the group symmetry is recovered in the limit $k\rightarrow\infty$.

Therefore, with a break of SU(3), our model displays composite operators
that are potential candidates for observable on the remaining group
structure, in close analogy with replica model \cite{Sorella:2010it}.
Note that the operators presented here may be useful in understanding
the spectroscopy of glueballs presented, eg in \cite{glueballsphysics},
since Gribov approach has shown promising results in this direction,
see the recent literature \cite{Dudal:2011Glueballmasses} and references
therein. This analysis is very complex and requires further investigation.

Here is important to comment that a dynamical symmetry breaking could
be the answer to a BRST invariante observable. In this case the full
BRST operator carries not only the gauge fields but also the auxiliary
fields. In this case it could be possible to define a BRST and group
invariant colorless physical observable.

\section{Conclusions}

In this work we have studied the $SU(3)$ Yang-Mills theory in a Landau
gauge with a soft mass term proportional to the symmetric structure
constant. This soft mass term alow us to treat the Gribov problem
in a local renormalizable action without the need of auxiliary fields
like in the Gribov-Zwanziger theory. The confining behavior is thus
induced by the soft mass term and the relation with the scaling solution
is presented in close analogy to the Gribov mechanism and to the replica
model\cite{Sorella:2010it}. Also the introduction of the diagonal
mass term and the relation with the decoupling solution is discussed.
Moreover this can open the possibility of a more general relation
between confining behavior and the existence of gauge condensates
in the infrared regime. In fact, it would be interesting the possibility
of obtaining the value of the mass gap by the local composite operator
method and study the relation between the extremun of the effective
potential and the Gribov mass gap equation.

The result we have obtained suggests further that the restriction
to the first Gribov region could be implemented into other gauges
by a simple soft mass term involving the symmetric structure constant
that breaks the original group into a more simple group struture with
a replica. Further studies aimed at establishing such a connection,
in particular for the maximal abelian gauge, in a more precise way
is called for and is a topic of current investigation.

Another important point analysed is the physical operators associated
to these propagator. Due to the structure of the propagator this is
a difficult task. The simple \textit{i-particle} structure is not
present in these model and the mechanism associated to the \textit{i-particle}
is much more complex and involves the group structure itself. This
is the price to be paid in order to do not double the number of gauge
fields or introduce auxiliary localizing fields. Nevertheless we find
possible candidates of physical operators associated to this propagator
in which the group structure is present in the ultraviolet regime.
It is argued that a dynamical breacking mechanism could restore the
full BRST invariance of the observable, in particular for $F_{\mu\nu}^{a}(x)F_{\mu\nu}^{a}(x)$
in the target remaining group structure. The construction of all this
mechanism for a target remaining group like an $SU(3)$ replica is
another topic to be investigated into future works.


\section*{Acknowledgments}


The Conselho Nacional de Desenvolvimento Científico e Tecnológico
(CNPq-Brazil), the FAPERJ, Funda{ç}{ã}o de Amparo {à} Pesquisa
do Estado do Rio de Janeiro, the SR2-UERJ and the Coordena{ç}{ã}o
de Aperfei{ç}oamento de Pessoal de N{í}vel Superior (CAPES) are
gratefully acknowledged for financial support.


\appendix

\section{Renormalizability}

\subsection{Equations Compatible with the Quantum Action Principle}

The full set of equations compatible with the quantum action principle
\cite{Piguet:1995er} is given by: 
\begin{itemize}
\item The Lagrange multiplier and the antighost equation: 
\begin{eqnarray}
\frac{\delta\Sigma}{\delta b^{a}} & = & i\partial_{\mu}A_{\mu}^{a}\nonumber \\
\frac{\delta\Sigma}{\delta\overline{c}^{a}}+\partial_{\mu}\frac{\delta\Sigma}{\delta\Omega_{\mu}^{a}} & = & 0\,.\label{lagrange}
\end{eqnarray}

\item The integrated ghost equation: 
\begin{eqnarray}
{\cal {G}}^{a}(\Sigma) & = & \Delta^{a}\nonumber \\
{\cal {G}}^{a}(\Sigma) & = & \int d^{4}x\lbrace\frac{\delta\Sigma}{\delta c^{a}}-igf^{abc}(\overline{c}^{b}\frac{\delta\Sigma}{\delta b^{c}}+\lambda^{b}\frac{\delta\Sigma}{\delta j^{c}})\rbrace,\nonumber \\
\Delta^{a} & = & \int d^{4}x\lbrace gf^{abc}(\Omega_{\mu}^{b}+\alpha\partial_{\mu}\lambda^{b})A_{\mu}^{c}-gf^{abc}(L^{b}c^{c}+i\varepsilon\lambda^{b}j^{c})\rbrace\,.\label{ghosteq}
\end{eqnarray}
Before presenting all the equations compatible with the quantum action
principle it is relevant to note here that the term $\frac{\varepsilon}{2}j^{a}j^{a}$
generates a linear breaking in the ghost equation. This will gave
us the information that this term does not renormalizes. 
\item Slavnov-Taylor: 
\begin{eqnarray}
{\cal {S}}(\Sigma) & = & \int d^{4}x\lbrace\frac{\delta\Sigma}{\delta\Omega_{\mu}^{a}}\frac{\delta\Sigma}{\delta A_{\mu}^{a}}+\frac{\delta\Sigma}{\delta L^{a}}\frac{\delta\Sigma}{\delta c^{a}}+ib^{a}\frac{\delta\Sigma}{\delta\overline{c}^{a}}+ij^{a}\frac{\delta\Sigma}{\delta\lambda^{a}}\rbrace\nonumber \\
{\cal {S}}(\Sigma) & = & 0\,.\label{slavnov}
\end{eqnarray}

\item Rigid symmetry: 
\begin{eqnarray}
W^{a}(\Sigma) & = & \int d^{4}xgf^{abc}\lbrace A_{\mu}^{b}\frac{\delta\Sigma}{\delta A_{\mu}^{c}}+c^{b}\frac{\delta\Sigma}{\delta c^{c}}+b^{b}\frac{\delta\Sigma}{\delta b^{c}}+\overline{c}^{b}\frac{\delta\Sigma}{\delta\overline{c}^{c}}+\Omega_{\mu}^{b}\frac{\delta\Sigma}{\delta\Omega_{\mu}^{c}}+L^{b}\frac{\delta\Sigma}{\delta L^{c}}+\lambda^{b}\frac{\delta\Sigma}{\delta\lambda^{c}}+j^{b}\frac{\delta\Sigma}{\delta j^{c}}\rbrace\nonumber \\
W^{a}(\Sigma) & = & 0\,.\label{rigid}
\end{eqnarray}

\item $SL(2,R)$: 
\begin{equation}
R(\Sigma)=\int d^{4}x\lbrace c^{a}\frac{\delta\Sigma}{\delta\overline{c}^{a}}-i\frac{\delta\Sigma}{\delta b^{ca}}\frac{\delta\Sigma}{\delta L^{a}}\rbrace=0\,.\label{sl2r}
\end{equation}

\end{itemize}

\subsection{Stability of the quantum action}

The next step is to characterize the most general counterterm that
can be freely added to all orders in perturbation theory respecting
all the symmetries presented previously. Following the set up of the
Algebraic Renormalization \cite{Piguet:1995er}, we perturb the classical
action $\Sigma$ by adding an integrated local polynomial in the fields
and sources, $\Sigma_{count}$, with dimension bounded by four, and
with vanishing ghost number. The perturbed action ($\Sigma+\epsilon\Sigma_{count}$),
where $\epsilon$ is an expansion parameter, fulfills, to the first
order in $\epsilon$, the same Ward identities obeyed by the classical
action $\Sigma$, \textit{i.e} equations (\ref{lagrange})-(\ref{sl2r}).
\begin{eqnarray}
\beta_{\Sigma}\Sigma_{count} & = & 0,\hspace{1cm}\frac{\delta\Sigma_{count}}{\delta b^{a}}=0,\hspace{1cm}(\frac{\delta}{\delta\overline{c}^{a}}+\partial_{\mu}\frac{\delta}{\delta\Omega_{\mu}^{a}})\Sigma_{count}=0,\nonumber \\
{\cal {G}}^{a}\Sigma_{count} & = & 0,\hspace{1cm}W^{a}\Sigma_{count}=0,\hspace{1cm}R\Sigma_{count}=0,\label{constraints}
\end{eqnarray}
where $\beta_{\Sigma}$ is given by: 
\begin{equation}
\beta_{\Sigma}=\int{d^{4}x}\lbrace\frac{\delta\Sigma}{\delta\Omega_{\mu}^{a}}\frac{\delta}{\delta A_{\mu}^{a}}+\frac{\delta\Sigma}{\delta A_{\mu}^{a}}\frac{\delta}{\delta\Omega_{\mu}^{a}}+\frac{\delta\Sigma}{\delta L^{a}}\frac{\delta}{\delta c^{a}}+\frac{\delta\Sigma}{\delta c^{a}}\frac{\delta}{\delta L^{a}}+ib^{a}\frac{\delta}{\delta\overline{c}^{a}}+ij^{a}\frac{\delta}{\delta\lambda^{a}}\rbrace\label{linearizado}
\end{equation}
thus, taking into account the general results on the cohomology of
Yang-Mills theories, the most general invariant counterterm is: 
\begin{equation}
\Sigma_{count}=\int{d^{4}x}\lbrace\frac{a_{0}}{4}F_{\mu\nu}^{a}F_{\mu\nu}^{a}\rbrace+\beta_{\Sigma}\Delta^{-1}.
\end{equation}
The integrated term corresponds to the nontrivial part of the cohomology
of $\beta_{\Sigma}$, while $\Delta^{-1}$ is an integrated polynomial
in the fields and sources with ultraviolet dimension 4 and ghost number
-1. The ultraviolet dimension and ghost number of all fields and sources
are presented in Table 1 below.

\begin{table}[h]
\centering %
\begin{tabular}{|c|c|c|c|c|c|c|c|c|}
\hline 
fields and sources  & $A_{\mu}^{a}$  & $c^{a}$  & $\overline{c}^{a}$  & $b^{a}$  & $\lambda^{a}$  & $j^{a}$  & $\Omega_{\mu}^{a}$  & $L^{a}$ \tabularnewline
\hline 
UV dimension  & 1  & 0  & 2  & 2  & 2  & 2  & 3  & 4 \tabularnewline
\hline 
Ghost number  & 0  & 1  & -1  & 0  & -1  & 0  & -1  & -2 \tabularnewline
\hline 
\end{tabular}\caption{Quantum numbers of fields and sources.}

\label{table1} 
\end{table}

Applying all the constraints given in (\ref{constraints}) and observing
the ultraviolet dimension $4$ and ghost number $-1$, we obtain for
$\Delta^{-1}$: 
\begin{eqnarray}
\Delta^{-1} & = & \int{d^{4}x}\lbrace a_{1}(\Omega_{\mu}^{a}+\partial_{\mu}\overline{c}^{a})A_{\mu}^{a}+a_{2}\alpha\partial_{\mu}\lambda^{a}A_{\mu}^{a}+\frac{a_{3}}{2}\lambda^{a}d^{abc}A_{\mu}^{b}A_{\mu}^{c}\rbrace
\end{eqnarray}

We see thus that $\Sigma_{count}$ contains $4$ free independent
parameters, namely $(a_{0},a_{1},a_{2},a_{3})$. These parameters
can be reabsorbed by means of a multiplicative renormalization of
the gauge coupling constant $g$, off the parameters $(\alpha,\varepsilon)$
and the set of fields and sources $\phi=(A_{\mu}^{a},c^{a},\overline{c}^{a},b^{a},\Omega_{\mu}^{a},L^{a},\lambda^{a},j^{a})$
according to 
\begin{equation}
\Sigma(g,\alpha,\phi)+\hbar\Sigma^{\mathrm{count}}=\Sigma(g_{0},\alpha_{0},\phi_{0})+O(\hbar^{2})\;,\label{reab}
\end{equation}
with 
\begin{eqnarray}
g_{0} & = & Z_{g}g,\hspace{0.3cm}\alpha_{0}=Z_{\alpha}\alpha,\hspace{0.3cm}\varepsilon_{0}=Z_{\varepsilon}\varepsilon\nonumber \\
A_{0\mu}^{a} & = & Z_{A}^{\frac{1}{2}}A_{\mu}^{a},\hspace{0.3cm}c_{0}^{a}=Z_{c}c_{0}^{a},\hspace{0.3cm}b_{0}^{a}=Z_{b}b^{a},\hspace{0.3cm}\overline{c}_{0}^{a}=Z_{\overline{c}}\overline{c}_{0}^{a},\nonumber \\
\lambda_{0}^{a} & = & Z_{\lambda}\lambda_{0}^{a},\hspace{0.3cm}j_{0}^{a}=Z_{j}j^{a},
\end{eqnarray}
and 
\begin{eqnarray}
Z_{g} & = & 1-\hbar\frac{a_{0}}{2},\hspace{0.3cm}Z_{\alpha}=1+\hbar(\frac{a_{0}}{2}-a_{2}-a_{3}),\hspace{0.3cm}Z_{\varepsilon}=1+\hbar(2a_{0}-2a_{3}),\nonumber \\
Z_{A}^{\frac{1}{2}} & = & 1+\frac{\hbar}{2}(a_{0}+2a_{1}),\hspace{0.3cm}Z_{c}=Z_{\overline{c}}=Z_{\Omega}=1-\hbar\frac{a_{1}}{2},\hspace{0.3cm}Z_{b}=1-\frac{\hbar}{2}(a_{0}+2a_{1}),\hspace{0.3cm}\nonumber \\
Z_{\lambda} & = & 1-\hbar(\frac{a_{0}}{2}-\frac{a_{1}}{2}-a_{3}),\hspace{0.3cm}Z_{j}=1-\hbar(a_{0}-a_{3}),\label{zes}
\end{eqnarray}
or directly in terms of the multiplicative relations between the renormalization
factors $Z$ 
\begin{equation}
Z_{b}Z_{A}^{\frac{1}{2}}=1,\hspace{0.5cm}Z_{\overline{c}}Z_{g}Z_{A}^{\frac{1}{2}}Z_{c}=1,\hspace{0.5cm}Z_{\varepsilon}Z_{j}^{2}=1.
\end{equation}

In order to clarify the importance of such relations and the Gribov
type propagator, let us follow the Zwanziger prescription and set
the sources $\lambda^{a},j^{a}$ respectively to $0,<j^{a}>$ with
$<j^{a}>\neq0$. Taking this into account it is clear that the action
is BRST invariant up to a soft breacking term proportional to $<j^{a}>$
\begin{eqnarray}
 &  & s\Sigma(A_{\mu}^{a},\overline{c}^{a},c^{a},b^{a},\lambda^{a}=0,L^{a}=0,\Omega_{\mu}^{a}=0,j^{a}=<j^{a}>)=<j^{a}>\Delta^{a}\nonumber \\
\Delta^{a} & = & \int{d^{4}x}\lbrace d^{abc}\partial_{\mu}c^{b}A_{\mu}^{c}-\frac{g}{2}f^{abc}c^{b}d^{cde}A_{\mu}^{d}A_{\mu}^{e}+\alpha\partial_{\mu}D_{\mu}^{ab}c^{b}\rbrace,
\end{eqnarray}
just like in the Zwanziger procedure to the Gribov problem and due
to the fact that we dont have localizing fields, the limit $<j^{a}>\rightarrow0$
clearly recover the pure Yang-Mills in the deep ultraviolet region.

\subsection{Nonrenormalization of the soft mass term, simple one loop prove}

Following closely the arguments presented by Sorella in \cite{Sorella:2010it}
we will prove that there is no one loop correction to the soft mass
term. The argument is based on dimensional regularization with minimal
subtraction and the fact that, at least at one loop, the tadpole diagram
in the two point $A-A$ is related to an integral of the type. 
\begin{equation}
\int{\frac{d^{D}k}{(2\pi)^{D}}}f^{adc}f^{bec}<A_{\mu}^{d}(k)A_{\nu}^{e}(k)>,\hspace{0.5cm}D=4-\epsilon\label{integral}
\end{equation}
which can be rewritten as 
\begin{eqnarray}
\int{\frac{d^{D}k}{(2\pi)^{D}}}\{f^{adc}f^{bdc}(\frac{k^{2}}{k^{4}+\frac{m^{4}}{4}}) & - & if^{adc}f^{bec}d^{de3}m^{2}(\frac{1}{k^{4}+\frac{m^{4}}{4}})\nonumber \\
 & + & f^{adc}f^{bec}\frac{m^{4}}{4(k^{4}+\frac{m^{4}}{4})}((\frac{1}{k^{2}})\sum_{i=1}^{2}\delta^{di}\delta^{ei}-\frac{1}{3}(\frac{k^{2}}{k^{4}+\frac{m^{4}}{3}})(\delta^{d8}\delta^{e8}+\delta^{d3}\delta^{e3})\nonumber \\
 & + & i\frac{m^{2}}{\sqrt{3}}\frac{1}{3}(\frac{1}{k^{4}+\frac{m^{4}}{3}})(\delta^{d8}\delta^{e3}+\delta^{d3}\delta^{e8}))\}
\end{eqnarray}

Now in order to prove that all terms do not contribute to the mass,
at least in first order of perturbation theory, let us analyse the
different terms of this integral. The first term give rise to: 
\begin{equation}
\int{\frac{d^{D}k}{(2\pi)^{D}}}\frac{k^{2}}{k^{4}+\mu^{4}}=\int{\frac{d^{D}k}{(2\pi)^{D}}}\frac{1}{k^{2}}-\int{\frac{d^{D}k}{(2\pi)^{D}}}\frac{\mu^{4}}{k^{2}(k^{4}+\mu^{4})},\label{separa}
\end{equation}
corresponding respectively to an integral that is zero by dimensional
and a power counting ultraviolet convergent one. The second term to
be analysed is of the form: 
\begin{equation}
\int{\frac{d^{D}k}{(2\pi)^{D}}}\frac{1}{k^{4}+\mu^{4}}=\int{\frac{d^{D}k}{(2\pi)^{D}}}\frac{1}{k^{4}}-\int{\frac{d^{D}k}{(2\pi)^{D}}}\frac{\mu^{4}}{k^{4}(k^{4}+\mu^{4})}.\label{separa2}
\end{equation}
Recursively the same argument is applied to the third integral, showing
thus that no divergent terms proportional to $<j^{3}>$ arise. Also,
from the absence of one-loop counterterm of the kind $<j^{a}>\frac{1}{2}d^{abc}A_{\mu}^{b}A_{\mu}^{c}$,
it follows that 
\begin{equation}
<j^{a}>_{0}\frac{1}{2}d^{abc}(A_{\mu}^{b})_{0}(A_{\mu}^{c})_{0}=<j^{a}>\frac{1}{2}d^{abc}A_{\mu}^{b}A_{\mu}^{c},
\end{equation}
so that we obtain 
\begin{equation}
<j^{a}>_{0}=Z_{<j>}<j^{a}>,\hspace{0.5cm}Z_{<j>}Z_{A}=1,
\end{equation}
meaning that the renormalization factor of the soft parameter $<j^{a}>$
can be expressed in terms of the gluon renormalization factor $Z_{A}$.
\footnote{A purely algebraic proof, valid to all orders, of the non-renormalization
properties of the soft parameter is under investigation. Also the
possibility of obtaining the value of $<j^{a}>$, by the local composite
operator method\cite{Dudal:2003vv,Dudal:2002aj}.%
}

\section{Taking a closer look at the $SU(3)$ group}

\label{su3}

\subsection{General considerations}

The $SU(N)$ group is the group of the $N\times N$ unitary matrices
with determinant equals to one: 
\begin{equation}
SU(N):=\{\, U\,\,|\,\, UU^{\dag}=\mathbf{1}\,,\,\det(U)=1\,\}\,.
\end{equation}
The matrices $U\in SU(N)$ can be written as 
\begin{equation}
U(\omega):=\exp\left(i\omega^{a}T^{a}\right)
\end{equation}
where $\omega^{a}$ is a parameter, the label $a$ runs from $1$
to $(N^{2}-1)$, and $T^{a}$ are the generators of the group, obeying
the following relations: 
\begin{eqnarray}
[T^{a},T^{b}] & = & if^{abc}T^{c}\,,\nonumber \\
\{T^{a},T^{b}\} & = & \frac{1}{N}\,\delta^{ab}+d^{abc}T^{c}\,.\label{algebra}
\end{eqnarray}
In equation (\ref{algebra}) $[\,\,,\,]$ stands for the commutator,
while $\{\,\,,\,\}$ for the anti-commutator; $f^{abc}$ are structure
constants, which are anti-symmetric by odd successive permutations,
\textit{i.e.} 
\begin{equation}
f^{abc}=-f^{bac}=-f^{acb}=-f^{cba}\,;
\end{equation}
and $d^{abc}$ are the components of the completely symmetric invariant
rank-3 tensor of the group.


\subsection{The $SU(2)$ group}

Before discuss the $SU(3)$ case it is useful to spend a few words
on the $SU(2)$ case. In this case there are three generators and
they are related with the Pauli matrices, $\sigma^{a}$, as follows:
\begin{equation}
T^{a}=\frac{\sigma^{a}}{2}\,,\qquad(a=1,2,3)\,,
\end{equation}
where, 
\begin{equation}
\sigma^{1}=\begin{pmatrix}0\end{pmatrix}\,,\qquad\sigma^{2}=\begin{pmatrix}0\end{pmatrix}\,,\qquad\sigma^{1}=\begin{pmatrix}1\end{pmatrix}\,.
\end{equation}
Also, we have 
\begin{equation}
f^{abc}=\varepsilon^{abc}\,,\qquad d^{abc}=0\,,
\end{equation}
and than 
\begin{equation}
\left[\,\frac{\sigma^{a}}{2}\,,\,\frac{\sigma^{b}}{2}\,\right]=i\varepsilon^{abc}\,\frac{\sigma^{c}}{2}\,,\qquad\left\{ \,\frac{\sigma^{a}}{2}\,,\,\frac{\sigma^{b}}{2}\,\right\} =\frac{1}{2}\,\delta^{ab}\,.
\end{equation}


\subsection{The $SU(3)$ group}

In the $SU(3)$ group there are eight generators associated with the
Gell-Mann matrices, $\lambda^{a}$, in an analog way as the Pauli
matrices for $SU(2)$, 
\begin{equation}
T^{a}=\frac{\lambda^{a}}{2}\,,\qquad(a=1,\dots,8)\,.
\end{equation}
The Lie algebra of the generators of $SU(3)$ is than given by 
\begin{equation}
\left[\,\frac{\lambda^{a}}{2}\,,\,\frac{\lambda^{b}}{2}\,\right]=if^{abc}\,\frac{\lambda^{c}}{2}\,,\qquad\left\{ \,\frac{\lambda^{a}}{2}\,,\,\frac{\lambda^{b}}{2}\,\right\} =\frac{1}{3}\,\delta^{ab}+d^{abc}\,\frac{\lambda^{c}}{2}\,.
\end{equation}
Here, the generators obey the anti-symmetric and symmetric Jacobi
identities: 
\begin{eqnarray*}
[\lambda^{a},[\lambda^{b},\lambda^{c}]]+[\lambda^{c},[\lambda^{a},\lambda^{b}]]+[\lambda^{b},[\lambda^{c},\lambda^{a}]] & = & 0\,,\\
{}[\lambda^{a},\{\lambda^{b},\lambda^{c}\}]+[\lambda^{c},\{\lambda^{a},\lambda^{b}\}]+[\lambda^{b},\{\lambda^{c},\lambda^{a}\}] & = & 0\,.
\end{eqnarray*}
These identities give rise to the following useful relations: 
\begin{eqnarray}
f^{abc}f^{cde}+f^{adc}f^{ceb}+f^{aec}f^{cbd} & = & 0\,,\nonumber \\
f^{ade}d^{dbc}+f^{cde}d^{dab}+f^{bde}d^{dca} & = & 0\,.\label{Ref: jacobi}
\end{eqnarray}
In particular, the second equation of (\ref{Ref: jacobi}) is satisfied
for $SU(N\geq3)$ and the first one is valid for any value of $N$.
Another interesting feature of the $SU(3)$ group is that its generators
can be grouped in sets which obeys the same algebraic properties of
the Pauli matrices determining then three $SU(2)$ subalgebras. Defining
\begin{equation}
\lambda_{\pm}=\frac{1}{2}(\sqrt{3}\,\lambda_{8}\pm\lambda_{3}),
\end{equation}
the three $SU(2)$ groups embedded in the $SU(3)$ are given by: 
\begin{eqnarray}
SU(2)_{\mathrm{I}} & : & \left(\frac{\lambda_{1}}{2},\frac{\lambda_{2}}{2},\frac{\lambda_{3}}{2}\right)\,,\nonumber \\
SU(2)_{\mathrm{II}} & : & \left(\frac{\lambda_{4}}{2},\frac{\lambda_{5}}{2},\frac{\lambda_{+}}{2}\right)\,,\nonumber \\
SU(2)_{\mathrm{III}} & : & \left(\frac{\lambda_{6}}{2},\frac{\lambda_{7}}{2},\frac{\lambda_{-}}{2}\right)\,.\label{3xSU(3)}
\end{eqnarray}


\subsubsection{The Gell-Mann matrices and the structure constants of $SU(3)$}

The Gell-Mann matrices are given by 
\begin{eqnarray}
 &  & \lambda_{1}=\begin{pmatrix}0\end{pmatrix}\,,\quad\lambda_{2}=\begin{pmatrix}0\end{pmatrix}\,,\quad\lambda_{3}=\begin{pmatrix}1\end{pmatrix}\,,\nonumber \\
 &  & \lambda_{4}=\begin{pmatrix}0\end{pmatrix}\,,\quad\lambda_{5}=\begin{pmatrix}0\end{pmatrix}\,,\quad\lambda_{6}=\begin{pmatrix}0\end{pmatrix}\,,\nonumber \\
 &  & \lambda_{7}=\begin{pmatrix}0\end{pmatrix}\,,\quad\lambda_{8}=\frac{1}{\sqrt{3}}\begin{pmatrix}1\end{pmatrix}\,.\quad
\end{eqnarray}
The nonzero structure constants are: 
\begin{eqnarray}
f^{123}=1\,,\qquad f^{147}=-f^{156}=f^{246}=f^{257}=f^{345}=-f^{367}=\frac{1}{2}\,,\qquad f^{458}=f^{678}=\frac{\sqrt{3}}{2}\,.
\end{eqnarray}
The nonzero components of the symmetric tensor $d^{abc}$ are: 
\begin{eqnarray}
 &  & d^{118}=d^{228}=d^{338}=-d^{888}=\frac{1}{\sqrt{3}}\,,\nonumber \\
 &  & d^{448}=d^{558}=d^{668}=d^{778}=-\frac{1}{2\sqrt{3}}\,,\nonumber \\
 &  & d^{146}=d^{157}=-d^{247}=d^{256}=d^{344}=d^{355}=-d^{366}=-d^{377}=\frac{1}{2}\,.
\end{eqnarray}

\end{document}